\documentclass{llncs}
\usepackage{amssymb,amsfonts,amstext,amsmath}
\usepackage{graphicx}
\usepackage{hyperref}
\hypersetup{
   colorlinks=true,
   urlcolor=blue,    
   linkcolor=black,  
   citecolor=blue,   
}
\begin{document}
\title{Refutability as Recursive as Provability} 
\titlerunning{Refutability as Recursive as Provability}
\author{Paola Cattabriga} 
\institute{University of Bologna \\  \medskip
\href{https://orcid.org/0000-0001-5260-2677}{ 0000-0001-5260-2677}}
\thispagestyle{empty}
 \maketitle
\bigskip
\begin{abstract}  
G\"{o}del numbering is an {\it arithmetization} of sintax which defines provability by coding a primitive recursive predicate, $\mathtt{Pf}$$(x,v)$  
\cite[190-198]{mendel}. A multiplicity of researches and results all around
 this well-known recursive predicate are today widespread in many areas of logic and AI. Not equally investigated is the refutability predicate defined by G\"{o}del numbering within the same primitive recursive status.
 $\mathtt{Rf}\mathrm (x,v)$ can be defined as a recursive predicate  meaning that $x$ 
is the G\"{o}del number of a refutation in 
{\it PA} of the formula with G\"{o}del number $v $.
This article proposes a logical investigation of the interactive links between provability and refutability predicates when defined within the same recursive status. The resulting Lemmas are clarifying and open new perspectives for the  incompleteness argument and the codings of its underlying notions. 
\end{abstract}

\section*{Introduction}
 Once the axioms and rules of inference in the calculus of first-order predicates have been established, thanks to G\"{o}del's 1930 completeness, all the theorems can be derived.  
 There are several formal ways to further describe deductive completeness, for example we can list, enumerate or generate all first-order theorems. Predicate calculus is considered semidecidable, since in first-order axiomatic theories we have a computable procedure for listing all predicate formulas that are theorems. We could also think of a machine, like the Turing machine, which writes all the first order theorems step by step on the tape.  Despite all this, there are arguments, such as the so-called Church and Turing theorems, which affirm the undecidability of first-order calculus.  We have truth tables as a decision method for propositional calculus, but we have no equivalent decision method for predicate calculus.  In simple terms, taking any first order formula we cannot always decide whether  it is true or false. At the basis of these arguments there is an upstream idea that sees the existence of formulas that are neither true nor false as an impasse. See for example the well-known words
 \begin{quote}
... there are propositions which are neither true nor false but {\it indeterminate} \cite[126] {luk22}.
\end{quote} 
A conception which in Lukasiewicz then seems to move from semantic validity to syntactic demonstrability and stated in another form:
 ``... there exist significant expressions, which can neither be proved by our axioms and rules of assertion nor disproved by our axioms and rules of rejection,  I call such expressions undecidable with respect to our basis" \cite[100] {luk51}.
This {\it indeterminacy} is  intrinsically linked to the non-denumerability and the infinite generation of formulas.
\begin{quote}
Is the number of undecidable expressions finite or not? If it is finite, the problem of decision is easily solved: we may accept true expressions as new asserted axioms, and reject false expressions axiomatically. The method, however, is not practicable if the number of undecidable expressions is not finite. We cannot assert or reject an infinity of axioms \cite[100] {luk51}.
\end{quote}
Something quite similar seems to echo from G\"{o}del. The completeness theorem establishes an equivalence between validity and provability, but this equivalence 
\begin{quote}
entails, for the decision problem, a reduction of the nondenumerable to the denumerable, since valid refers to the nondenumerable totality of functions, while provable presupposes only the denumerable totality of the formal proofs \cite{godel30}.
\end{quote}
From a historical point of view, we can trace a conceptual path that leads to {\it indeterminacy}  in the  face of infinity, going back to the Lowenheim-Skolem theorem, the Skolem paradox and  the so-called Cantor theorem about the non-denumerability of the set of all the subsets of the natural numbers \cite{skolem,cantor1,cantor2}. As is known, diagonalization is the basis of Cantor's reasoning and consequently also of Skolem's paradox.  
Moreover, G\"{o}del's  1931 undecidable formula itself contains the codification of a form of self-referentiality, usually called the  Diagonalization Lemma \cite{godel1}\cite[362-363]{godel2}\cite[203]{mendel}. In this regard we have already pointed out the  neglect and omission of the complement of the set or predicate made object of diagonalization \cite{catta3,catta4}.  
G\"{o}del derives the undecidable formula encoding the notion of provability by means of the famous primitive recursive predicate $\mathtt{Pf}$$(x,v)$  
\cite[190-198]{mendel}. This derivation is achieved by completely neglecting the underlying definability of an equally recursive encoding of refutability \cite{catta1,catta2,catta5}. As noted by Lukasiewicz:
\begin{quote}
Of two intellectual acts, to assert a proposition and to reject it, only the first has been taken into account in modern formal logic \cite[95] {luk51}.
\end{quote}
Indeterminacy leads Lukasiewicz not only to reject the principle of bivalence and to introduce a third truth value, but also to the need to formalize the rejection.
This aligns with our exploration of the definition of the complementary predicate of $\mathtt{Pf}$$(x,v)$, namely $\mathtt{Rf}\mathrm (x,v)$ \cite{catta1,catta5}. 
Below we reproduce and extend the investigation of the relations between these two predicates, since it seems to provide another way out of  \emph{indeterminacy}.

\bigskip

\subsection*{Basic Setup}
We shall assume a first order theory which adequately formalizes 
Peano Arithmetic (see 
for example the system $S$, with all the necessary assumptions, in 
\cite{mendel} 116-175).
 Let us call it {\it PA}. 
  Numerals, as usual,
are  defined recursively, $\overline{0}$ is $0$ and for any natural 
number $n$, $\overline{n+1}$
is $(\overline{n})'$ (where $'$ is the Successor function). For any 
expression $X$  we use $\ulcorner X \urcorner$ to denote the 
corresponding
G\"{o}del number of $X$. Let us define the G\"{o}del numbering as 
follows:
\begin{enumerate}
\item First assign different odd numbers to the primitive symbols of 
the language
of {\it PA}.
\item Let $X$ be a formal expression $X_{0},X_{1},\dots,X_{n}$, where 
each $X_i$, 
$0\leqslant i\leqslant n$, is
a primitive symbol of the language of  {\it PA}. Then
$$ \ulcorner X\urcorner\,=\,p_{0}^{\ulcorner 
X_{0}\urcorner}\centerdot 
p_{1}^{\ulcorner X_{1}\urcorner}\centerdot\ldots\centerdot 
p_{n}^{\ulcorner X_{n}\urcorner}$$
where $p_n$ is the $n$-th prime number and $p_{0}=2$.
\item Let $X$ be composed by the formal expressions 
$X_{0},X_{1},\dots,X_{n}$, then
$$ \ulcorner X\urcorner\,=\,p_{0}^{\ulcorner 
X_{0}\urcorner}\centerdot 
\ldots\centerdot p_{n}^{\ulcorner X_{n}\urcorner}.$$
\end{enumerate}
For any given  formula  $\phi(v)$ of {\it PA} we then have  its 
G\"{o}del number 
$n\,=\,\ulcorner\phi (v)\urcorner$. This number $n$ has a name in the 
language
 of {\it PA},
namely $\overline n$, and this name can be substituted back into 
$\phi(v)$. This
self-reference procedure is usually admitted by the so-called 
Diagonalization Lemma \cite{catta1,catta3,catta4}.

\section{Refutation Predicate}\label{R}
We shall construct a new predicate by  G\"{o}del numbering. 
The reader can refer to the arithmetization as defined by Mendelson;  
the new predicate must be considered as a last relations added 
to the functions and  relations  (1-26) presented 
in \cite[190-198]{mendel}. We shall not reproduce entirely 
this 
long list of definitions which is already  well-known (see also 
\cite[162-176]{godel1}).  
Let us start  recalling  some of the definitions  involved, precisely 
only
those we need.
    
$\mathtt{MP}$$(x,y,z)$: The expression  with  G\"{o}del number $z$ is 
a direct consequence
of the expressions with G\"{o}del numbers $x$ and $y$ by modus ponens,
$$y = 2^{3} * x * 2^{11} * z * 2^{5} \wedge\mathtt{Gd}
\mathrm (x)\wedge\mathtt{Gd}\mathrm (z). $$
  
$\mathtt{Gen}$$(x,y)$: The expression  with  G\"{o}del number $y$ 
comes from
 the expression with G\"{o}del number $x$  by the Generalization Rule,
 
\smallskip

\noindent $(\exists v)_{v<y} (\mathtt{EVbl}\mathrm (v) \wedge 
y = 2^{3} *  2^{3} *  2^{13} * v * 2^{5} * x * 2^{5} \wedge 
\mathtt{Gd}\mathrm (x)). $
  
\smallskip
  
$\mathtt{Ax}$$(y)$: $y$ is the   G\"{o}del number of an axiom of {\it 
PA}:
$$\mathtt{LAx}\mathrm (y)\vee\mathtt{PrAx}\mathrm (y). $$
  
$\mathtt{Neg}$$(v)$: the G\"{o}del number of $(\neg \alpha)$ if $v$ 
is the  G\"{o}del 
number of $ \alpha$:
$$ \mathtt{Neg}\mathrm (v) = 2^{3} * 2^{9} * v * 2^{5}.$$
  
$\mathtt{Prf}$$(x)$: $x$ is the G\"{o}del number of a proof in {\it 
PA}: 

\smallskip

\noindent $
\exists u_{u<x}\; \exists v_{v<x} \;\exists z_{z<x}\; \exists 
w_{w<x}\; ([x  =2^{w} \wedge \mathtt{Ax}\mathrm (w)] \vee $

\noindent $
[\mathtt{Prf}\mathrm (u) \wedge  \mathtt{Fml}\mathrm ((u)_{w})  \wedge  x  = u * 2^{v}  
\wedge \mathtt{Gen}\mathrm ((u)_{w},v)]\vee $ 

\noindent$
[\mathtt{Prf}\mathrm (u) \wedge \mathtt{Fml}\mathrm ((u)_{z}) \wedge  
\mathtt{Fml}\mathrm ((u)_{w})  \wedge   x  = u * 2^{v} \wedge 
 \mathtt{MP}\mathrm ((u)_{z},(u)_{w},v)]\vee $

\noindent $[\mathtt{Prf}\mathrm (u) \wedge  x  = u * 2^{v} \wedge 
\mathtt{Ax}\mathrm (v)]).
$
  
\smallskip
  
$\mathtt{Pf}$$(x,v)$: $x$ is the G\"{o}del number of a proof in {\it 
PA} of the
 formula with G\"{o}del number $v$: $$ \mathtt{Prf}\mathrm (x) \wedge 
v = (x)_{\mathit{lh}
 \mathrm (x) \overset{\centerdot}{\text{--}} 1}.$$

By means of such definitions, we shall define a new 
predicate,  $\mathtt{Rf}$.

\bigskip
 
$\mathtt{Rf}\mathrm (x,v)$: $x$ is the G\"{o}del number of a proof in 
{\it PA} of the negation of the formula with G\"{o}del 
number $v$: $$ \mathtt{Pf} \mathrm (x,z) \wedge z = \mathtt{Neg}\mathrm (v). $$

\bigskip

\noindent In other terms  $\mathtt{Rf} \mathrm (x,v)$ states $x$ 
is the G\"{o}del number of a refutation in 
{\it PA} of the formula with G\"{o}del number $v$.
$\mathtt{Rf}$ is \emph{primitive recursive}, as the 
relations 
obtained from primitive recursive relations by means of 
propositional connectives are also primitive recursive \cite[69]{hermes}\cite[180]{mendel}.
For their recursiveness  $\mathtt{Pf} \mathrm (x,v)$ and $\mathtt{Rf} \mathrm (x,v)$  are
\emph{expressible} in {\it PA} by the formulas $Pf(x,v)$ and $Rf(x,v)$.  
Hereafter we will call  $Pf(x,v)$ and $Rf(x,v)$ respectively the \emph{proof} and the \emph{refutation} predicate, while ``provability" and ``refutability" will refer to their existential assertions (section \ref{bility}).

\section{Proof and Refutation connections}\label{PR}
From Luckasievicz's `assertion - rejection' binomial   to our current `proof - refutation' we can see a historical progression, which goes from those systems of demonstration traditionally on the level of the thought and pure symbolic representation up to the systems of effective computation and automated deduction in computer science and artificial intelligence.  Automatic theorem proving branch from two traditions of logical inquiry, the first  via \emph{resolution} from Herbrand's theorem and Robinson's resolution rule, and the second via \emph{tableaux} from Gentzen's natural deduction. The mechanical generation of first order theorems can be considered as an investigation that starts precisely from the deductive completeness of predicate calculus. The recursive definition of the proof predicate is historically the result  of investigations both on
 the search of an algorithm to solve any mathematical problem and on
 problems of proof theory.
The classical provability problem is that to decide, for a given a first order formula, 
 if it is provable in the formal system  \cite{bgg}.
For its recursiveness the proof predicate $\mathtt{Pf}(x,y)$ must be understood as an algorithm, i.e. an effective computation procedure, which outputs  those first order arithmetical formulas that are the theorems of  {\it PA}. The sequence of 45 numbered formulas defining the proof predicate 
in G\"{o}del's 1931 looks very much like a computer program, so that Kurt G\"{o}del's appears today to be like a pioneer of logic programming  \cite[97-98]{davis1}\cite[14-19]{chaitin}.

To visualize the algorithmic structure of this predicates  let us display two simple examples. The first one, a propositional logic case, where $x $ is the G\"{o}del number of a proof of a formula with G\"{o}del number  $ y $. 

\begin{example}\label{esempio1}
Let us suppose we have the proof of  $\vdash \neg \neg B \longrightarrow B$ as follows.					
\[
\begin{array}[t]{ll}
(1)
& \quad \quad ( \neg B \longrightarrow \neg \neg B)  \longrightarrow   (( \neg B \longrightarrow \neg B) \longrightarrow B),  \\
(2) 
 &\quad \quad ( \neg B \longrightarrow \neg B),\\
(3) 
 &\quad \quad ( \neg B \longrightarrow \neg\neg B) \longrightarrow B,\\
(4) 
 &\quad \quad \neg \neg B \longrightarrow ( \neg B \longrightarrow \neg\neg B),\\
 (5) 
 &\quad \quad \neg \neg B \longrightarrow  B.
\end{array}\] 
The corresponding  G\"{o}del numbers for each (1) - (5)  would be:
\[
\begin{array}[t]{ll}
g(\mathtt{\boldsymbol{1}} )
& \quad \quad 2^{3} 3^{9} 5^{51} 7^{11} 11^{9} 13^{9} 17^{51} 19^{5}  23^{11} 29^{3} 31^{3} 37^{9} 41^{51} 43^{11} 47^{9} 53^{51} 59^{5} 61^{11} 67^{51} 71^{5}  \\ 
g(\mathtt{\boldsymbol{2}} ) 
 &\quad \quad 2^{3} 3^{9} 5^{51} 7^{11} 11^{9} 13^{51} 17^{5}\\ 
g(\mathtt{\boldsymbol{3}} )
 &\quad \quad 2^{3} 3^{9} 5^{51} 7^{11} 11^{9} 13^{9} 17^{51} 19^{5}  23^{11} 29^{5} \\ 
g(\mathtt{\boldsymbol{4}} )
 &\quad \quad      2^{9} 3^{9} 5^{51} 7^{11} 11^{3} 13^{9} 17^{51} 19^{11} 23^{9} 29^{9} 31^{51} 37^{5} \\ 
g(\mathtt{\boldsymbol{5}} ) 
 &\quad \quad 2^{3} 3^{9} 5^{9} 7^{51} 11^{11} 13^{51} 17^{5}
\end{array}\] 
So that in $\mathtt{Pf}(x,y)$ $x$  would be an expression like 
$\; 2^{g(\mathtt{\boldsymbol{1}} )} u \,  3^{g(\mathtt{\boldsymbol{2}} )}  u \,  5^{g(\mathtt{\boldsymbol{3}} )}  u  \,7^{g(\mathtt{\boldsymbol{4}} )}  u \,11^{g(\mathtt{\boldsymbol{5}} )}\; $  and $ y = g(\mathtt{\boldsymbol{5}} )$, (i.e. $ \overline{\ulcorner\neg \neg B \longrightarrow B\urcorner} $).
\end{example}
\begin{example}\label{esempio2}	
As a further simple example, the derivation in eleven steps by $\mathtt{Rf}(x,y)$ of  $\; \vdash \neg (t < t) \;$ in formal number 
theory:
\[
\begin{array}[t]{llll}
(1) &  \quad  t < t,  &\quad (6) & \quad  b + t = 0 + t, \\
(2) & \quad  \exists w ( w \neq 0 \wedge  w + t = t), &\quad (7) &\quad b = 0, \\
(3) & \quad   b \neq 0 \wedge  b + t = t, &\quad (8)  &\quad  b \neq 0, \\
(4) & \quad b + t = t, &\quad (9)  &\quad  b = 0 \wedge b \neq 0,\\
(5)  & \quad  t = 0 + t, &\quad (10) &\quad  0 = 0 \wedge 0 \neq 0,\\ 
 & & \quad (11) &\quad \neg ( t < t)\\
 \end{array}\] 
\cite[164]{mendel}.
It would yield  for $x$  an expression like
$\; u \, 2^{g(\mathtt{\boldsymbol{1}} )} u \,  3^{g(\mathtt{\boldsymbol{2}} )}  \dots $ $ \, 23^{g(\mathtt{\boldsymbol{10}} )}  u \, 29^{g(\mathtt{\boldsymbol{11}} )}\; $  and for $y$  $ \; 2^{33}  \,  3^{865}   \,5^{33 } \; $ (i.e. $\overline{\ulcorner t < t \urcorner}$)\footnote{ We regarded $t$ as to be $f^0_1$ and $< $ as $f^2_3$.}.
\end{example}
The link from each formula to its G\"{o}del number is a one to one correspondence ensured by  the uniqueness of the factorization of the integers into primes \cite[191]{mendel}. To every primitive recursive predicate there corresponds  a primitive recursive  characteristic function, and since every primitive recursive function is computable then every recursive predicate is  decidable \cite[66-67]{hermes}\cite[51-55]{davis}.
Accordingly, both the proof predicate $\mathtt{Pf}(x,y)$ and the refutation predicate $\mathtt{Rf}(x,y)$ are also decidable. The first  says us whether $x$ is the G\"{o}del number of a proof of the formula with G\"{o}del number  $y$ or not, while the second says us whether $x$ is the G\"{o}del number of a refutation of the formula with G\"{o}del number  $y$ or not.
 It therefore seems legitimate to ask, given such effective procedures, are there actually undecidable formulas? Or, in other words, where exactly is \emph{indeterminacy} encoded? The connections between the two recursive predicates produces the followings lemmas,  that will be useful for answering this questions (see section \ref{bility}).

\bigskip

\begin{lemma}\label{notboth}
For any natural number $n$ and for any formula $\alpha$  not both
$\mathtt{Rf}\mathrm (n,\ulcorner \alpha \urcorner)$ 
and $\mathtt{Pf}\mathrm (n,\ulcorner \alpha \urcorner)$.
\end{lemma}
\begin{proof}
Let us suppose to have both $\mathtt{Rf}\mathrm (n,\ulcorner \alpha 
\urcorner)$ and
$\mathtt{Pf}\mathrm (n,\ulcorner \alpha \urcorner)$.
We should have then
$\mathtt{Prf}\mathrm (n) \wedge \ulcorner \alpha \urcorner = 
(n)_{\mathit{lh}
 \mathrm (n) \overset{\centerdot}{\text{--}} 1}$ and $\mathtt{Pf} 
\mathrm (n,z) \wedge 
z = \mathtt{Neg}\mathrm (\ulcorner \alpha \urcorner),$ 
i.e. 
$\mathtt{Prf}\mathrm (n) \wedge \ulcorner \alpha \urcorner = 
(n)_{\mathit{lh}
 \mathrm (n) \overset{\centerdot}{\text{--}} 1}$
 and  $\mathtt{Prf} \mathrm (n) \wedge 
\mathtt{Neg}\mathrm (\ulcorner \alpha \urcorner) = (n)_{\mathit{lh}
 \mathrm (n) \overset{\centerdot}{\text{--}} 1}.$

By the definition of $\mathtt{Prf} \mathrm (x) $ this would mean to 
have

\noindent $
\exists u_{u<n}\; \exists v_{v<n} \;\exists z_{z<n}\; \exists 
w_{w<n}\; ([n=2^{w} \wedge \mathtt{Ax}\mathrm (w)] \vee $

\noindent $[\mathtt{Prf}\mathrm (u) \wedge \mathtt{Fml}\mathrm ((u)_{w}) \wedge 
n = u * 2^{v} \wedge \mathtt{Gen}\mathrm ((u)_{w},v)]\vee $

\noindent $[\mathtt{Prf}\mathrm (u) \wedge \mathtt{Fml}\mathrm ((u)_{z}) \wedge 
\mathtt{Fml}\mathrm ((u)_{w})  \wedge n = u * 2^{v} \wedge 
\mathtt{MP}\mathrm ((u)_{z},(u)_{w},v)]\vee $

\noindent $[\mathtt{Prf}\mathrm (u) \wedge  n = u * 2^{v} \wedge 
\mathtt{Ax}\mathrm (v)]
$ 
and both
$$\mathrm \ulcorner \alpha \urcorner = (n)_{\mathit{lh}
 \mathrm (n) \overset{\centerdot}{\text{--}} 1}
\:\:\text{ and }\:\:
\mathtt{Neg}\mathrm (\ulcorner \alpha \urcorner) = (n)_{\mathit{lh}
 \mathrm (n) \overset{\centerdot}{\text{--}} 1}
$$
and hence the four cases
\begin{enumerate}
	\item  $[n=2^{\ulcorner \alpha \urcorner} \wedge \mathtt{Ax}\mathrm 
(\ulcorner \alpha \urcorner)] 
\text{ and }$

\noindent $ [n=2^{\mathtt{Neg}\mathrm (\ulcorner \alpha \urcorner) } 
\wedge \mathtt{Ax}\mathrm (\mathtt{Neg}\mathrm (\ulcorner \alpha 
\urcorner) )] $

	\item  \noindent $[\mathtt{Prf}\mathrm (u) \wedge \mathtt{Fml}\mathrm ((u)_{w}) 
\wedge n = u * 2^{\ulcorner \alpha \urcorner} \wedge 
\mathtt{Gen}\mathrm ((u)_{w},\ulcorner \alpha \urcorner)]
 \text{ and }$
 
 \noindent $  [\mathtt{Prf}\mathrm (u) \wedge \mathtt{Fml}\mathrm 
((u)_{w}) \wedge n = u * 2^{\mathtt{Neg}\mathrm (\ulcorner \alpha 
\urcorner) } \wedge \mathtt{Gen}\mathrm ((u)_{w},\mathtt{Neg}\mathrm 
(\ulcorner \alpha \urcorner) )] $

	\item  $[\mathtt{Prf}\mathrm (u) \wedge \mathtt{Fml}\mathrm ((u)_{z}) 
\wedge \mathtt{Fml}\mathrm ((u)_{w})  \wedge n = u * 2^{\ulcorner 
\alpha \urcorner} \wedge \mathtt{MP}\mathrm 
((u)_{z},(u)_{w},\ulcorner \alpha \urcorner)]
 \text{ and }$
 
 \noindent $
[\mathtt{Prf}\mathrm (u) \wedge \mathtt{Fml}\mathrm ((u)_{z}) \wedge 
\mathtt{Fml}\mathrm ((u)_{w})  \wedge n = u * 2^{\mathtt{Neg}\mathrm 
(\ulcorner \alpha \urcorner)} \wedge \mathtt{MP}\mathrm 
((u)_{z},(u)_{w},\mathtt{Neg}\mathrm (\ulcorner \alpha \urcorner))] $

	\item  $[\mathtt{Prf}\mathrm (u) \wedge  n = u * 2^{\ulcorner \alpha 
\urcorner} \wedge \mathtt{Ax}\mathrm (\ulcorner \alpha \urcorner)]
\text{ and }$

\noindent $  [\mathtt{Prf}\mathrm (u) \wedge  n = u * 
2^{\mathtt{Neg}\mathrm (\ulcorner \alpha \urcorner) } \wedge 
\mathtt{Ax}\mathrm (\mathtt{Neg}\mathrm (\ulcorner \alpha \urcorner) 
)]
$ 
\end{enumerate}
which are all immediately impossible  by the definitions of 
$\mathtt{Ax}\mathrm (y)$,
$\mathtt{Gen}\mathrm (x,y)$ and $\mathtt{MP}\mathrm (x,y,z)$  and 
thence 
by the  definitions of
 the axioms of {\it PA},    the Generalization 
Rule and  Modus Ponens, because no axiom  belongs to {\it PA} 
together with its 
negation and  the two inference rules    preserve
logical validity. 
\end{proof}

We now recall  the definition of characteristic function. If $R$ is a
relation of $n$ arguments, then the characteristic function $C_R$ is 
defined as follows
\begin{equation}\notag
 C_R(x_1,\dots,x_n) =
\begin{cases}
0& \text{if $R(x_1,\dots,x_n)$ is true,}\\
1& \text{if $R(x_1,\dots,x_n)$ is false.}
\end{cases} 
\end{equation} 

Let us call the characteristic functions of $\mathtt{Pf}\mathrm 
(x,v)$ and 
 $\mathtt{Rf}\mathrm (x,v)$ 
respectively $C_{\mathtt{Pf}}$ and
$C_{\mathtt{Rf}}$.
A relation $R(x_1,\dots,x_n)$ is said to be primitive recursive 
(recursive) if and only if its
characteristic function $C_R(x_1,\dots,x_n)$ is primitive recursive 
(recursive) \cite[179-180]{mendel} .
As $\mathtt{Pf}\mathrm (x,v)$ and   
$\mathtt{Rf}\mathrm (x,v)$  are  
primitive recursive 
then also $C_{\mathtt{Pf}}$ and $C_{\mathtt{Rf}}$  are  primitive recursive.
Every recursive function is representable in {\it PA}, thence
 $C_{\mathtt{Pf}}$  and
 $C_{\mathtt{Rf}}$  are
representable in {\it PA}.
We shall  assume   $C_{Pf}$ and 
 $C_{Rf}$  to represent respectively 
 $C_{\mathtt{Pf}}$ and
 $C_{\mathtt{Rf}}$  in {\it PA}.
\begin{lemma}\label{complete1}
For any formula $\alpha$, and $n$ as the   G\"{o}del number of a proof
in {\it PA} of $\alpha$
$$ 
\vdash _{PA}\: C_{Pf}(\overline{n},\overline{\ulcorner 
\alpha\urcorner})=\overline{0}
 \;\;\wedge\;\; 
C_{Rf}(\overline{n},\overline{\ulcorner \alpha\urcorner})=\overline{1}
$$
\end {lemma} 
\begin{proof}
One can easily see  that the two conjuncts are true: as $n$ is the   
G\"{o}del number of a proof
in {\it PA} of $\alpha$ 
$C_{Pf}(\overline{n},\overline{\ulcorner 
\alpha\urcorner})=\overline{0}$ is true. 
By Lemma (\ref{notboth}) $\mathtt{Rf}\mathrm(n,\ulcorner 
\alpha\urcorner)$ does not hold,
therefore it is true that
 $n$ is not the   G\"{o}del number of a refutation
in {\it PA} of $\alpha$.  
\end{proof}
\begin{lemma}\label{complete2}
For any formula $\alpha$, and $n$ as the   G\"{o}del number of a 
refutation
in {\it PA} of $\alpha$
$$ 
\vdash _{PA}\: C_{Rf}(\overline{n},\overline{\ulcorner 
\alpha\urcorner})=\overline{0}
 \;\;\wedge\;\; 
C_{Pf}(\overline{n},\overline{\ulcorner \alpha\urcorner})=\overline{1}
$$
\end {lemma} 
\begin{proof}
One can easily see  that the two conjuncts are true: as $n$ is the   
G\"{o}del 
number of a refutation
in {\it PA} of $\alpha$ 
$C_{Rf}(\overline{n},\overline{\ulcorner 
\alpha\urcorner})=\overline{0}$ is true. 
By Lemma (\ref{notboth}) $\mathtt{Pf}\mathrm(n,\ulcorner 
\alpha\urcorner)$ does not hold,
therefore it is true that
 $n$ is not the   G\"{o}del number of a proof
in {\it PA} of $\alpha$.  
\end{proof}
\begin{lemma}\label{antidiag1}
For  any formula $\alpha$ 
 
\begin{itemize}
	\item[(I)]\it{ not both }  $$\vdash _{PA}\:  Pf(\overline{n},\overline{\ulcorner 
\alpha\urcorner}) \:\: \vdash _{PA} \: Rf(\overline{n},\overline{\ulcorner 
\alpha\urcorner}),$$
    \item[(II)] \it{for } $n$  \it{ as the   G\"{o}del number of a refutation
in  PA of } $\alpha$ $$ \vdash _{PA} \: Rf(\overline{n},\overline{\ulcorner \alpha\urcorner}) \Longleftrightarrow 
\neg Pf(\overline{n},\overline{\ulcorner \alpha\urcorner}),$$
	\item[(III)]\it{for } $n$  \it{ as the   G\"{o}del number of a proof 
	in  PA of } $\alpha  $
$$\vdash _{PA} \:  Pf(\overline{n},\overline{\ulcorner \alpha\urcorner}) 
\Longleftrightarrow \neg Rf(\overline{n},\overline{\ulcorner 
\alpha\urcorner}).$$
\end{itemize}
\end{lemma}
\begin{proof}
(I) Immediately by Lemma (\ref{notboth}) and the definition of being 
expressible which holds for 
both $\mathtt{Pf}\mathrm (x,v)$ and  $\mathtt{Rf}\mathrm (x,v)$ 
(\cite{mendel} 130).

(II) Let us assume $\vdash _{PA} \: 
Rf(\overline{n},\overline{\ulcorner \alpha\urcorner})$, then
Lemma (\ref{complete2}) yields $\vdash _{PA} \: 
C_{Pf}(\overline{n},\overline{\ulcorner 
\alpha\urcorner})=\overline{1}$.
Hence by definition  $Pf(\overline{n},\overline{\ulcorner 
\alpha\urcorner})$ is false,
consequently $\vdash _{PA} \: \neg 
Pf(\overline{n},\overline{\ulcorner \alpha\urcorner})$.
Conversely let us assume 
$\vdash _{PA} \: \neg Pf(\overline{n},\overline{\ulcorner 
\alpha\urcorner})$
then $Pf(\overline{n},\overline{\ulcorner \alpha\urcorner})$ is false 
and by Lemma
(\ref{complete2}) we attain $\vdash _{PA} \: 
Rf(\overline{n},\overline{\ulcorner \alpha\urcorner})$.

(III) Let us assume $\vdash _{PA} \: 
Pf(\overline{n},\overline{\ulcorner \alpha\urcorner})$, then
Lemma (\ref{complete1}) yields $\vdash _{PA} \: 
C_{Rf}(\overline{n},\overline{\ulcorner 
\alpha\urcorner})=\overline{1}$.
Hence by definition  $Rf(\overline{n},\overline{\ulcorner 
\alpha\urcorner})$ is false,
consequently $\vdash _{PA} \: \neg 
Rf(\overline{n},\overline{\ulcorner \alpha\urcorner})$.
Conversely let us assume 
$\vdash _{PA} \: \neg Rf(\overline{n},\overline{\ulcorner 
\alpha\urcorner})$
then $Rf(\overline{n},\overline{\ulcorner \alpha\urcorner})$ is false 
and by Lemma
(\ref{complete1}) we attain 
$\vdash _{PA} \: Pf(\overline{n},\overline{\ulcorner 
\alpha\urcorner})$.
\end{proof}

\section{Provability and Refutability}\label{bility}
In general, an n-ary predicate $P$ is recursively enumerable if  it is empty or if there exist $n$ singularly recursive functions $f_1\dots f_n$ such that for all $x_1\dots x_n$  $$P(x_1\dots x_n) \iff \exists y ( f_1(y) = x_1 \wedge \dots \wedge f_n(y) = x_n ).$$ 
In addiction, every predicate $\exists y P( x_1\dots x_n y )$ with a recursive kernel  $P$ is a recursively enumerable predicate, r.e. predicate, and every r.e. predicate can be represented in this way \cite[188-189]{hermes} \cite[103]{borger}. In other terms, applying an unlimited existential quantifier to a predicate whose kernel is recursive yields a r. e. predicate. This is the case of the famous ``$\text{Bew}(x)$'',  
the first one of the list of notions 
1--46 in G\"{o}del's 1931 of which we cannot say that it is recursive, being defined as an assertion of existence. Let us quote its definition  \cite[171]{godel1}:

\bigskip

\begin{quote}

\begin{math}  
\text{45. } xBy \equiv Bw(x)\;\&\; [l(x)]\, Gl\, x = y, \\
\text{$x$ is a PROOF of the FORMULA y.}
\end{math}
\end{quote}

\begin{quote}
\begin{math}  
\text{46. } \text{Bew}(x) \equiv (Ey) y B x ,\\
\text{$x$ is a PROVABLE FORMULA. }
\end{math}.

 \end{quote}
 
\smallskip

$(Ey) y B x$  states it does exist a proof $y$ of the formula $x$, and  it is accordingly assumed throughout  G\"{o}del's 1931 argumentation as an assertion of existence, to derive the famous undecidable formula.  Let us consider $\text{Bew}(x)$, i.e.  in our modern notation $\exists y Pf(y,x)$, which assumes the existence of a  proof $y$ of $x$, in the light of our  lemmas.  

 \smallskip

There is no doubt that from ``$n$  {\it as the G\"{o}del number of a proof in  PA of} $\alpha  $" and  $\exists y Pf(y,x)$ we obtain $Pf(\overline{n},\overline{\ulcorner \alpha\urcorner})$.
Accordingly by Lemma (\ref{antidiag1})(III) 
$\vdash _{PA} \:  Pf(\overline{n},\overline{\ulcorner \alpha\urcorner}) 
\Longleftrightarrow \neg Rf(\overline{n},\overline{\ulcorner 
\alpha\urcorner}).$
And, on the other hand, from
 ``  $n$  {\it as the G\"{o}del number of a refutation in  PA of} $\alpha  $" and  $\exists y Rf(y,x)$, we have  $Rf(\overline{n},\overline{\ulcorner \alpha\urcorner})$, and by Lemma (\ref{antidiag1})(II) 
$\vdash _{PA} \:  Rf(\overline{n},\overline{\ulcorner \alpha\urcorner}) 
\Longleftrightarrow \neg Pf(\overline{n},\overline{\ulcorner 
\alpha\urcorner}).$  

Let us firstly resume the usual way to reason about as follows.

\begin{itemize}
\item[a)]
Being $\mathtt{Pf}\mathrm (x,v)$ decidable, for $Pf(\overline{n},\overline{\ulcorner\alpha\urcorner})$  we can decide whether $n$ is the G\"{o}del number of a proof
in {\it PA} of $\alpha$, namely whether
  $ \vdash _{PA}  \alpha$ or  $ \nvdash _{PA}  \alpha$.  
But when  we assert $\exists x Pf(x,\overline{\ulcorner \alpha\urcorner})$ then we can only generate all the formulas $ \alpha$ such that $ \vdash _{PA} \alpha$, and we could not be able to generate all the formulas such that   $\nvdash _{PA} \alpha$.
\item[b)]
By $\mathtt{Rf}\mathrm (x,v)$, for $Rf(\overline{n},\overline{\ulcorner\alpha\urcorner})$  we can decide whether $n$ is the G\"{o}del number of a refutation
in {\it PA} of $\alpha$, so  we can decide whether  $\vdash _{PA} \neg \alpha$ or $\nvdash _{PA} \neg  \alpha$.
Likewise, to state $\exists x Rf(x,\overline{\ulcorner \alpha\urcorner})$  surely generate all the formulas $\alpha$ such that $ \vdash _{PA} \neg \alpha$, but we could not be able to generate all the formulas such that $\nvdash _{PA} \neg \alpha$.
\end{itemize}

But now thanks to Lemma (\ref{antidiag1})  when in PA we assume that $\exists x Pf(x,\overline{\ulcorner \alpha\urcorner})$  (or  $\exists x Rf(x,\overline{\ulcorner \alpha\urcorner})$ ) we know something more i.e.  (\ref{antidiag1})(III) (or (\ref{antidiag1})(II)).

\begin{itemize}
\item[a$^\prime$)]
If in PA we state that $\exists x Pf(x,\overline{\ulcorner \alpha\urcorner})$ we can add (\ref{antidiag1})(III), then for  $n$   as the   G\"{o}del number of a proof
in  PA of  $\alpha$, we have not only $ \vdash _{PA} \:  Pf(\overline{n},\overline{\ulcorner \alpha\urcorner}) $ and hence $ \vdash _{PA} \alpha$, but also $\vdash _{PA} \neg Rf(\overline{n},\overline{\ulcorner 
\alpha\urcorner})$, namely $ \nvdash _{PA} \neg \alpha$.
\item[b$^\prime$)]
When in PA we state  $\exists x Rf(x,\overline{\ulcorner \alpha\urcorner})$ we can add (\ref{antidiag1})(II). That is 
for  $n$   as the   G\"{o}del number of a refutation
in  PA of  $\alpha$,   not only $ \vdash _{PA} \: Rf(\overline{n},\overline{\ulcorner \alpha\urcorner})$, i.e. $ \vdash _{PA} \neg \alpha$,
but consequently also 
$ \vdash _{PA} \:
\neg Pf(\overline{n},\overline{\ulcorner \alpha\urcorner}),$ namely $ \nvdash _{PA}  \alpha$.
\end{itemize}
This also explains our previous results in \cite{catta1,catta2,catta3,catta5}. 
\begin{remark} \rm
We are then able to notice that, being the predicate $\mathtt{Rf}\mathrm (x,v)$ recursively defined, as much as $\mathtt{Pf}\mathrm (x,v)$ is, it comes to our aid with respect to G\"{o}del's 1931  ``$ \text{Bew}(x) $". As noticed in the introduction this is something at all in twilights within G\"{o}del's incompleteness argument.
Moreover, the recursive definition of $\mathtt{Rf}\mathrm (x,v)$ highlights how, apart from the self-referring nested in the definition $Q(x,y)$ \cite[175]{godel1}, the assumption of the existence of a proof $y$ when stating $(Ey) y B x$  is precisely a codification, or formal expression, of the \emph{indeterminacy}. Indeterminacy, that in the light of the Lemma (\ref{antidiag1}) has no real reason to exist.
\end{remark}
As a concluding paragraph, let us explain further  how all the four possible cases $ \vdash\alpha$, $ \nvdash  \alpha$ and  $ \vdash \neg \alpha$ , $ \nvdash \neg \alpha$  are now ruled in {\it PA}.

As already mentioned,  $C_{Pf}$ is computable and $Pf(x,y)$ decidable. For $n$ a  G\"{o}del number of a proof
in {\it PA} of $\alpha$, $\mathbb{<}\overline{n},\overline{\ulcorner \alpha\urcorner} \mathbb{>}\in Pf \leftrightarrow C_{Pf}(\overline{n},\overline{\ulcorner \alpha\urcorner}) = \overline{0}$, and we can decide whether  $ \vdash _{PA}  \alpha$ or  $ \nvdash _{PA}  \alpha$.
See example \ref{esempio1}, $< 2^{g(\mathtt{\boldsymbol{1}} )} u \,  3^{g(\mathtt{\boldsymbol{2}} )}  u \,  5^{g(\mathtt{\boldsymbol{3}} )}  u $ $ \,7^{g(\mathtt{\boldsymbol{4}} )}  u \,11^{g(\mathtt{\boldsymbol{5}} )}, \overline{\ulcorner\neg \neg B \longrightarrow B\urcorner} > \in Pf$, where we can decide whether $\neg \neg B \longrightarrow B$ is a theorem or not, in  {\it PA}.
But when we state   $\exists x Pf(x,\overline{\ulcorner \alpha\urcorner})$ we make an assumption of existence, i.e.  of a r. e. predicate, that can be represented as  follows
$$\mathit{i)} \quad Pf(0,\overline{\ulcorner \alpha\urcorner})\vee Pf(1,\overline{\ulcorner \alpha\urcorner})\vee Pf(2,\overline{\ulcorner \alpha\urcorner})\vee Pf(3,\overline{\ulcorner \alpha\urcorner})\vee  \dots   $$
 \cite[xxiv-xxv]{davis}.
 In other terms,  we have a computable function $f$ which lists all the  G\"{o}del numbers of proofs
in {\it PA} yielding the set of all the ordered couples $\mathbb{<}f(0), \overline{\ulcorner \alpha\urcorner}\mathbb{>},  \mathbb{<}f(1),\overline{\ulcorner \alpha\urcorner}\mathbb{>},$ $ \mathbb{<}f(2),\overline{\ulcorner \alpha\urcorner}\mathbb{>} \dots $ till to reach that $\overline{n}$ such that effectively $\mathbb{<}\overline{n},\overline{\ulcorner \alpha\urcorner} \mathbb{>}\in Pf$. 
See the example \ref{esempio1},  then we would be restricted by $\mathit{i)}$ to have a computable function enumerating all the proof in {\it PA} till to reach the first couple such that 
$< 2^{g(\mathtt{\boldsymbol{1}} )} u \,  3^{g(\mathtt{\boldsymbol{2}} )}  u \,  5^{g(\mathtt{\boldsymbol{3}} )}  u $ $ \,7^{g(\mathtt{\boldsymbol{4}} )}  u \,11^{g(\mathtt{\boldsymbol{5}} )}, \overline{\ulcorner\neg \neg B \longrightarrow B\urcorner} > \in Pf$. Being $\exists x Pf(x,\overline{\ulcorner \alpha\urcorner})$  only r.e. we can generate the   G\"{o}del number of a proof
in {\it PA} of $\alpha$, i.e.
 $ \vdash _{PA} \alpha$, but we could not be able to obtain $ \nvdash _{PA} \alpha$. 
 
 \smallskip
 
 Respectively,  $C_{Rf}$ $\mathbb{<}\overline{n},\overline{\ulcorner \alpha\urcorner}\mathbb{>}\in Rf \leftrightarrow C_{Rf}(\overline{n},\overline{\ulcorner \alpha\urcorner}) = 0$, and since $Rf(x,y)$ is decidable we can decide whether  $n$  is the is the   G\"{o}del number of a refutation
in {\it PA} of $\alpha$, i.e. whether $ \vdash _{PA}  \neg\alpha$ or  $ \nvdash _{PA}  \neg\alpha$.  In example \ref{esempio2},
 we can decide whether $ t < t$ is a refutation or not in {\it PA}.
But  with the assertion  $\exists x Rf(x,\overline{\ulcorner \alpha\urcorner})$ we  have a r.e. predicate, such that
$$\mathit{ii)} \quad Rf(0,\overline{\ulcorner \alpha\urcorner})\vee Rf(1,\overline{\ulcorner \alpha\urcorner})\vee Rf(2,\overline{\ulcorner \alpha\urcorner})\vee Rf(3,\overline{\ulcorner \alpha\urcorner})\vee  \dots  \; ,$$
 and we can reach $ \vdash_{PA} \neg\alpha$, but we could not be able to generate a derivation such that $ \nvdash_{PA} \neg \alpha$. 
In example \ref{esempio2} we could be able to reach  $< u \,2^{g(\mathtt{\boldsymbol{1}} )} u \,  3^{g(\mathtt{\boldsymbol{2}} )} \dots u \,  23^{g(\mathtt{\boldsymbol{10}} )}  u $ 
 $ \,29^{g(\mathtt{\boldsymbol{11}} )},\overline{\ulcorner t < t  \urcorner}
 > \in Rf$ but we could never obtain a derivation such that $ \nvdash_{PA}  \neg(  t < t )$. 
 
 \bigskip
 
 Let us therefore  extend our investigation  for  the assertion of existence. For ``$n$  {\it  the G\"{o}del number of a proof in  PA of} $\alpha  $" from  $\exists x Pf(x,y)$ we obtain $Pf(\overline{n},\overline{\ulcorner \alpha\urcorner})$.
By Lemma (\ref{antidiag1})(III) 
$\vdash _{PA} \:  Pf(\overline{n},\overline{\ulcorner \alpha\urcorner}) 
\Longleftrightarrow \neg Rf(\overline{n},\overline{\ulcorner 
\alpha\urcorner})$, and from the viewpoint of the effective computability this is equivalent to generating from $\mathit{i)} \quad$ 
$$\mathit{iii)} \quad \neg Rf(0,\overline{\ulcorner \alpha\urcorner})\vee \neg Rf(1,\overline{\ulcorner \alpha\urcorner})\vee \neg Rf(2,\overline{\ulcorner \alpha\urcorner})\vee \neg Rf(3,\overline{\ulcorner \alpha\urcorner})\vee \dots \; .$$
In accordance, from $\mathit{i)}$ and $\mathit{iii)}$ we can attain both $ \vdash _{PA} \alpha$ and $ \nvdash _{PA} \neg \alpha$. 
On the other side, for ``$n$  {\it is the G\"{o}del number of a refutation in  PA of} $\alpha  $", $\exists x Rf(x,y)$,  we can obtain $Rf(\overline{n},\overline{\ulcorner \alpha\urcorner})$.
Then by Lemma (\ref{antidiag1})(II) 
$\vdash _{PA} \:  Rf(\overline{n},\overline{\ulcorner \alpha\urcorner}) 
\Longleftrightarrow \neg Pf(\overline{n},\overline{\ulcorner 
\alpha\urcorner})$, from $\mathit{ii)} \quad$ 
$$\mathit{iv)} \quad \neg Pf(0,\overline{\ulcorner \alpha\urcorner})\vee \neg Pf(1,\overline{\ulcorner \alpha\urcorner})\vee \neg Pf(2,\overline{\ulcorner \alpha\urcorner})\vee \neg Pf(3,\overline{\ulcorner \alpha\urcorner})\vee \dots \; ,$$
and from  $\mathit{ii)}$ and $\mathit{iv)}$ we can obtain both  $ \nvdash _{PA} \alpha$ and $ \vdash _{PA} \neg \alpha$.
Let us notice also that an assertion of existence like $\exists x \neg Pf(x,y)$ would on its turn generate the lists $\mathit{iv)}$ and $\mathit{ii)}$, while $\exists x \neg Rf(x,y)$ would yield $\mathit{iii)}$ and $\mathit{i)}$.   

By reason of all that, about provability and refutability,  \emph{indeterminacy} has no reason to be.  And this is also an explanation of the reason behind our previous results in \cite{catta1,catta2,catta3,catta5}.
 For objections about theories with more than denumerable models see \cite{catta3,catta4}.
 
\bigskip


\end{document}